\documentclass[prc,preprint,superscriptaddress,showpacs,nofootinbib,
tightenlines]{revtex4}
\usepackage{graphics}
\usepackage{epsfig}
\begin{document}
\preprint{MKPH-T-05-02}
\title{A new determination of the $\gamma \pi \rightarrow \pi\pi$ anomalous
amplitude via $\pi^- e^- \rightarrow \pi^- e^- \pi^{0}$ data}
\author{I.~Giller}
\altaffiliation{Present address:
Parametric Technology Israel Ltd.,
37 Havazelet Hasharon St., Beit Noy,
46641 Herzlia, Israel}
\affiliation{School of Physics and Astronomy,
Raymond and Beverly Sackler Faculty of Exact Sciences,
Tel Aviv University, 69978 Tel Aviv, Israel}
\author{A.~Ocherashvili}
\altaffiliation{Present address:
Advanced Technology Center, Sheba Medical Center, Tel Hashomer, 52621
Ramat Gan, Israel} \affiliation{School of Physics and Astronomy, 
Raymond and Beverly 
Sackler Faculty of Exact Sciences, Tel Aviv University, 69978 Tel Aviv, 
Israel}
\author{T.~Ebertsh\"auser}
\altaffiliation{Present address: Deutsche Forschungsgemeinschaft,
Kennedyallee 40, D-53175 Bonn, 
Germany}
\affiliation{Institut f\"ur Kernphysik, Johannes Gutenberg-Universit\"at,
J.~J.~Becher-Weg 45, D-55099 Mainz, Germany}
\author{M.~A.~Moinester}\email{murraym@tauphy.tau.ac.il}
\affiliation{School of Physics and Astronomy,
Raymond and Beverly Sackler Faculty of Exact Sciences,
Tel Aviv University, 69978 Tel Aviv, Israel}
\author{S.~Scherer}\email{scherer@kph.uni-mainz.de}
\affiliation{Institut f\"ur Kernphysik, Johannes Gutenberg-Universit\"at,
J.~J.~Becher-Weg 45, D-55099 Mainz, Germany}
\date{March 21, 2005}
\begin{abstract}
   We discuss the reaction $\pi^- e^- \rightarrow \pi^- e^- \pi^{0}$ with
the purpose of obtaining information on the  $\gamma \pi \rightarrow \pi\pi$
anomalous amplitude ${\cal F}_{3\pi}$.
   We compare a full calculation at ${\cal O}(p^6)$ in chiral perturbation
theory and various phenomenological predictions with the existing data of
Amendolia {\em et al}. 
   By integrating our theory results using Monte Carlo techniques we obtain 
$\sigma= 2.05 $ nb at ${\cal O}(p^6)$ and $\sigma= 2.17 $ nb after including
the dominant electromagnetic correction. 
   Both results are in good agreement with the experimental cross section of 
$\sigma= (2.11 \pm 0.47)$ nb.
   On the basis of the ChPT results one would extract from the 
the experimental cross section as amplitudes ${\cal F}_{3\pi}^{(0)\rm extr}
= (9.9 \pm 1.1)$ GeV$^{-3}$ and ${\cal F}_{3\pi}^{(0)\rm extr}
= (9.6 \pm 1.1)$ GeV$^{-3}$, respectively, 
which have to be compared with the low-energy theorem 
${\cal F}_{3\pi}=e/(4\pi^2 F_\pi^3)=9.72\,\mbox{GeV}^{-3}$. 
   We emphasize the need for new data to allow for a comparison of
experimental and theoretical distributions and to obtain ${\cal F}_{3\pi}$ 
with smaller uncertainty.
\end{abstract}
\maketitle

\section{\label{section_introduction}Introduction and Overview}
   Ever since the late 1960s anomalies
\cite{Adler:1969gk,Adler:1969er,Bardeen:1969md,Bell:1969ts}
have played an important role in our understanding of strong-interaction
physics.
   Anomalies arise if the symmetries of the Lagrangian at the classical
level are not supported by the quantized theory after renormalization,
resulting in so-called anomalous Ward identities \cite{Bardeen:1969md}.
   For the case of chiral $\mbox{SU(3)}_L\times\mbox{SU(3)}_R$, the
constraints due to the anomalous Ward identities have efficiently been
taken care of through the effective Wess-Zumino-Witten (WZW) action
\cite{Wess:yu,Witten:tw}.
   The WZW action is expressed in terms of the pseudoscalar octet of
Goldstone bosons and contributes at ${\cal O}(p^4)$ in the momentum expansion 
of chiral perturbation theory \cite{chpt} (for an overview see, e.g., Refs.\
\cite{chptreviews,Scherer:2002tk}).
   It is determined through the underlying group structure up to an overall
constant \cite{Wess:yu} and, in the purely strong sector, gives rise to
interaction vertices involving an odd number of Goldstone bosons
(odd-intrinsic-parity sector) \cite{Witten:tw}.
   Using topological arguments, Witten has shown that the WZW action is
quantized, i.e.\ a multiple of an integer parameter $n$.
   By including a coupling to electromagnetism, this parameter has been
identified as the number of colors $N_c$ by comparing with the prediction
of the QCD triangle diagram for the (anomalous) decay $\pi^0\to\gamma\gamma$.
   Once the overall factor is fixed, the (gauged) WZW action also predicts
other interactions such as the $\gamma\pi^+\pi^0\pi^-$ vertex.
   However, it has recently been pointed out by B\"ar and Wiese
\cite{Bar:2001qk} that the $N_c$ dependence in the pion-photon vertices
is completely canceled once the $N_c$ dependence of the quark charges
is consistently taken into account.
   In that sense, the width of the decay $\pi^0\to \gamma\gamma$ is predicted
absolutely without reference to the number of colors.
   The conclusion from their analysis is that one should rather consider
three-flavor processes such as $\eta\to\pi^+\pi^-\gamma$ or
$K\gamma\to K\pi$ to test the expected $N_c$ dependence
\cite{Bar:2001qk,Rogalyov:2002bg} in a low-energy reaction.
   However, by investigating the corresponding $\eta$ and $\eta'$
decays up to next-to-leading order in the framework of the combined $1/N_c$ and
chiral expansions, Borasoy and Lipartia have concluded that the number of 
colors cannot be determined from these decays due to the importance of 
sub-leading terms which are needed to account for the experimental decay 
widths and photon spectra \cite{Borasoy:2004mf}.

   The decay $\pi^{0}\rightarrow\gamma\gamma$ is the prime example of an
anomalous process \cite{Adler:1969gk,Bell:1969ts} and its invariant amplitude 
can be written as
\begin{equation}
\label{mpi0gg} {\cal M}_{\pi^0\to\gamma\gamma}= i {\cal F}_\pi(M_{\pi^0}^2)
\epsilon_{\mu\nu\rho\sigma} q_1^\mu\epsilon_1^{\nu \ast}
q_2^\rho\epsilon_2^{\sigma\ast},\quad \epsilon_{0123}=1.
\end{equation}
   The prediction in the chiral limit, as obtained from the WZW action,
is given by \cite{Wess:yu,Witten:tw,Bar:2001qk}
\begin{equation}
\label{calfpi0pred} {\cal F}_\pi(0)= \frac{\alpha}{\pi F_0},
\end{equation}
where $\alpha=e^2/4\pi\approx 1/137$, $e>0$,  and $F_0$ denotes the SU(3) 
chiral limit of the pion-decay constant \cite{chpt}: 
$F_\pi=F_0[1+{\cal O}(m_q)]=92.4\,
\mbox{MeV}$ \cite{Eidelman}.
   Using Eq.\ (\ref{calfpi0pred}) with the empirical value $F_\pi$ instead
of $F_0$, one obtains for the decay rate
\begin{equation}
\Gamma_{\pi^0\to\gamma\gamma}=\frac{\alpha^2 M_{\pi^0}^3}{64 \pi^3 F_\pi^2}
=7.73\, \mbox{eV}
\label{eq:int2}
\end{equation}
in agreement with the average value of Ref.\ \cite{Eidelman}:
\begin{equation}
\Gamma_{\pi^0\to\gamma\gamma}=(7.74\pm 0.55)\, \mbox{eV}.
\label{eq:int3}
\end{equation}
   Corrections due to explicit chiral symmetry breaking have been studied in
Refs.\ \cite{Donoghue:wv,Bijnens:1988kx,Moussallam:1994xp,%
Ananthanarayan:2002kj,Nasrallah:2002yi,Goity:2002nn}.
   The most recent analyses yield
$(8.06\pm0.02\pm 0.06)\,\mbox{eV}$ \cite{Ananthanarayan:2002kj} in
$\mbox{SU(2)}_L\times\mbox{SU(2)}_R$ chiral perturbation theory at ${\cal
O}(p^6)$ including electromagnetic corrections at ${\cal O}(e^2 p^4)$, 
$(8.60\pm 0.10)\,\mbox{eV}$ \cite{Nasrallah:2002yi} in the framework of a 
dispersion theory approach, and $(8.10\pm 0.08)\,\mbox{eV}$ 
\cite{Goity:2002nn} using $\mbox{U(3)}_L\times\mbox{U(3)}_3$ chiral 
perturbation theory at ${\cal O}(p^6)$ in combination with large-$N_c$ 
arguments.
   As has been stressed in Ref.\ \cite{Goity:2002nn}, the individual
experimental results show a large dispersion and depend on the specific
reaction used to extract the amplitude.
   The Primakoff Experiment at Jefferson Lab (PrimEx) \cite{PrimEx}
aims at a determination of the width with an accuracy of 1.5 \%
and will thus match the increased precision of the theoretical
analysis.

   As mentioned above, the WZW action also predicts more complicated
processes such as the $\gamma\pi^+\pi^0\pi^-$ interaction
and one clearly needs to confirm our picture of both the leading-order term
as well as the relevant corrections.
   The invariant amplitude for $\gamma^\ast(q)+\pi^-(p_b)\to
\pi^0(p_2)+\pi^-(p_3)$ can be written as
\begin{equation}
{\cal M}_{\gamma\pi^-\to\pi^0\pi^-}(q,p_b;p_2,p_3) =-i{\cal
F}_{3\pi}(s_2,t_2,u_2;q^2)\epsilon_{\mu\nu\rho\sigma} \epsilon^\mu
p_b^\nu p_2^\rho p_3^\sigma,
\end{equation}
where the Mandelstam variables are defined as $s_2=(q+p_b)^2$, 
$t_2=(p_b-p_3)^2$,
$u_2=(p_b-p_2)^2$ and satisfy the standard relation $s_2+u_2+t_2=2
M_{\pi^-}^2+M_{\pi^0}^2+q^2$.\footnote{Our notation and kinematics will be
discussed in more detail in Sec.\ \ref{section_kinematics}.}
   The lowest-order prediction [${\cal O}(p^4)$] is independent of $s_2$,
$t_2$, $u_2$, and $q^2$ \cite{Wess:yu,Witten:tw,Terent'ev:1971kt},
\begin{equation}
{\cal F}_{3\pi}=\frac{e}{4 \pi^{2}F^{3}_0} \approx \frac{e}{4
\pi^{2}F^{3}_\pi}=9.72\,\mbox{GeV}^{-3}. \label{eq:int5}
\end{equation}
   The physical threshold for $q^2=0$ is at 
$s_2^{\rm thr}=(M_{\pi^-}+M_{\pi^0})^2$,
$t_2^{\rm thr}=-M_{\pi^-}M_{\pi^0}^2/ (M_{\pi^-}+M_{\pi^0})$, 
and $u_2^{\rm thr}=
M_{\pi^-} (M_{\pi^-}^2-M_{\pi^-}M_{\pi^0}-M_{\pi^0}^2)/(M_{\pi^-}+M_{\pi^0})$.

   The amplitude ${\cal F}_{3\pi}$ was measured by Antipov {\em et al.}
\cite{Antipov:1986tp} at Serpukhov using 40 GeV pions.
   Their study involved pion pair production by pions in the nuclear Coulomb
field via the Primakoff reaction
\begin{equation}
{\pi^-} + (Z,A) \rightarrow {\pi^-}' + (Z,A) + {\pi^0}, \label{eq:bin2}
\end{equation}
where $Z$ and $A$ denote the nuclear charge and mass number, respectively.
   In the one-photon-exchange domain, Eq.\ (\ref{eq:bin2}) is equivalent to
\begin{equation}
 {\pi^-} + \gamma^\ast  \rightarrow  {\pi^-}' + {\pi^0}
\label{eq:bin3}
\end{equation}
with an almost real photon ($q^2\approx 0$).
   Diffractive production of the two-pion final state is blocked by $G$-parity
conservation.
   At CERN COMPASS \cite{Baum:1996yv}, a physics program based on pion and kaon
scattering from the nuclear Coulomb field (Primakoff scattering
\cite{Moinester:1997dm}) has begun.
   The program goals include state-of-the-art measurements of the chiral
anomaly transitions $\pi^-+\gamma^\ast\to{\pi^-}'+\pi^0$ 
and $K^-+\gamma^\ast\to
{K^-}'+\pi^0$ as well as measurements of pion and kaon polarizabilities and
radiative transitions \cite{Moinester:2003rb} and hybrid meson production
\cite{Moinester:2003ra}.

   The chiral anomaly sample of Ref.\ \cite{Antipov:1986tp} 
(roughly 200 events)
covered the ranges $s_2 < 10\, M_\pi^2$ and $|t_2| < 3.5\, M_\pi^2<s_2$.
   The small $t_2$ range selects events predominantly associated with the
exchange of a virtual photon, for which the target nucleus acts as a spectator.
   Assuming a constant amplitude ${\cal F}_{3\pi}$, the value
\begin{equation}
{\cal F}_{3\pi}^{\rm exp}=(12.9\pm 0.9\pm 0.5)\,\mbox{GeV}^{-3} \label{eq:int6}
\end{equation}
was extracted from the experiment \cite{Antipov:1986tp}.
   The considerable discrepancy with the theoretical prediction of
Eq.\ (\ref{eq:int5}) has generated a lot of interest in clarifying the
situation from both the experimental and theoretical sides.

   Higher-order corrections in the odd-intrinsic-parity sector of ChPT
have extensively been studied by Bijnens {\em et al.}
\cite{Bijnens:1988kx,Bijnens:1989jb,Bijnens:1989ff,Bijnens:xi}.
   They included one-loop diagrams involving one vertex from the WZW term and
tree diagrams from the anomalous ${\cal O}(p^6)$ Lagrangian 
\cite{anop6,Ebertshauser:2001nj,Bijnens:2001bb}, where
the parameters of the Lagrangian have been estimated via vector-meson dominance
(VMD) calculations.
   While the higher-order corrections to ${\cal F}_{\pi}$ of Eq.\
(\ref{calfpi0pred}) are small,
  for ${\cal F}_{3\pi}$ they increase the lowest-order value
between 7 \% and 12 \% within the kinematic range of the Serphukov experiment
\cite{Bijnens:1989ff}.
   Moreover, genuine one-loop corrections and ${\cal O}(p^6)$
tree-level contributions were found to be comparable in size.
   It has also been stressed by Holstein \cite{Holstein:1995qj} that the
experimental value of Eq.\ (\ref{eq:int6}) was obtained under the assumption 
of a {\em constant} amplitude whereas a re-analysis using a suitable 
dependence on the kinematical variables would produce a lower value 
\cite{Holstein:1995qj}
\begin{equation}
{\cal F}_{3\pi}^{\rm exp}=(11.9\pm 0.9\pm 0.5)\, \mbox{GeV}^{-3},
\label{eq:int6_1}
\end{equation}
and thus reduce the difference between theory and experiment.
   A sophisticated analysis has been carried out by Hannah \cite{Hannah:2001ee}
in the framework of a two-loop evaluation [${\cal O}(p^8)$] using dispersive
methods.
   From a comparison with the Antipov {\em et al.}\ data with the two-loop
analysis leaving ${\cal F}_{3\pi}^{(0)}$ as a free parameter,
Hannah obtained
\begin{equation}
{\cal F}_{3\pi}^{\rm exp}=(11.4\pm 1.3)\, \mbox{GeV}^{-3}. \label{eq:thtab}
\end{equation}
   By also including radiative corrections, Ametller {\em et al.}\
\cite{Ametller:2001yk} showed that the electromagnetic corrections
generate a sizeable increase in the Primakov cross section,
leading, in comparison with Eq.\ (\ref{eq:thtab}), to a further
decrease
\begin{equation}
{\cal F}_{3\pi}^{\rm exp}=(10.7\pm 1.2)\,\mbox{GeV}^{-3}. \label{eq:akt}
\end{equation}
   Finally, using an integral equation approach, Truong \cite{Truong:2001en}
obtained
\begin{equation}
{\cal F}_{3\pi}=11.2\,\mbox{GeV}^{-3}. \label{eq:truong}
\end{equation}
   Further theoretical investigations of ${\cal F}_{3\pi}$ include
calculations in the framework of dynamical constituent quarks \cite{dynquarks}.

   The limited accuracy of the existing data in combination with the various 
new calculations clearly motivates improved and more precise experiments
\cite{Baum:1996yv,Moinester:1994gh,Miskimen}.
   In a recent JLab experiment \cite{Miskimen},
results on $\gamma\to 3\pi$ were obtained from an analysis of $\gamma p\to
\pi^+\pi^0 n$ data taken with the CLAS detector.
   The photon energy was approximately 2 GeV.
   A Chew-Low analysis was used to extract ${\cal F}_{3\pi}$ from the cross
sections over a large kinematic range.
   Preliminary results were presented by B.~Asavapibhop \cite{Asavapibhop}
and an experimental paper is in preparation \cite{Miskimen:private}.

\begin{figure}
\epsfig{file=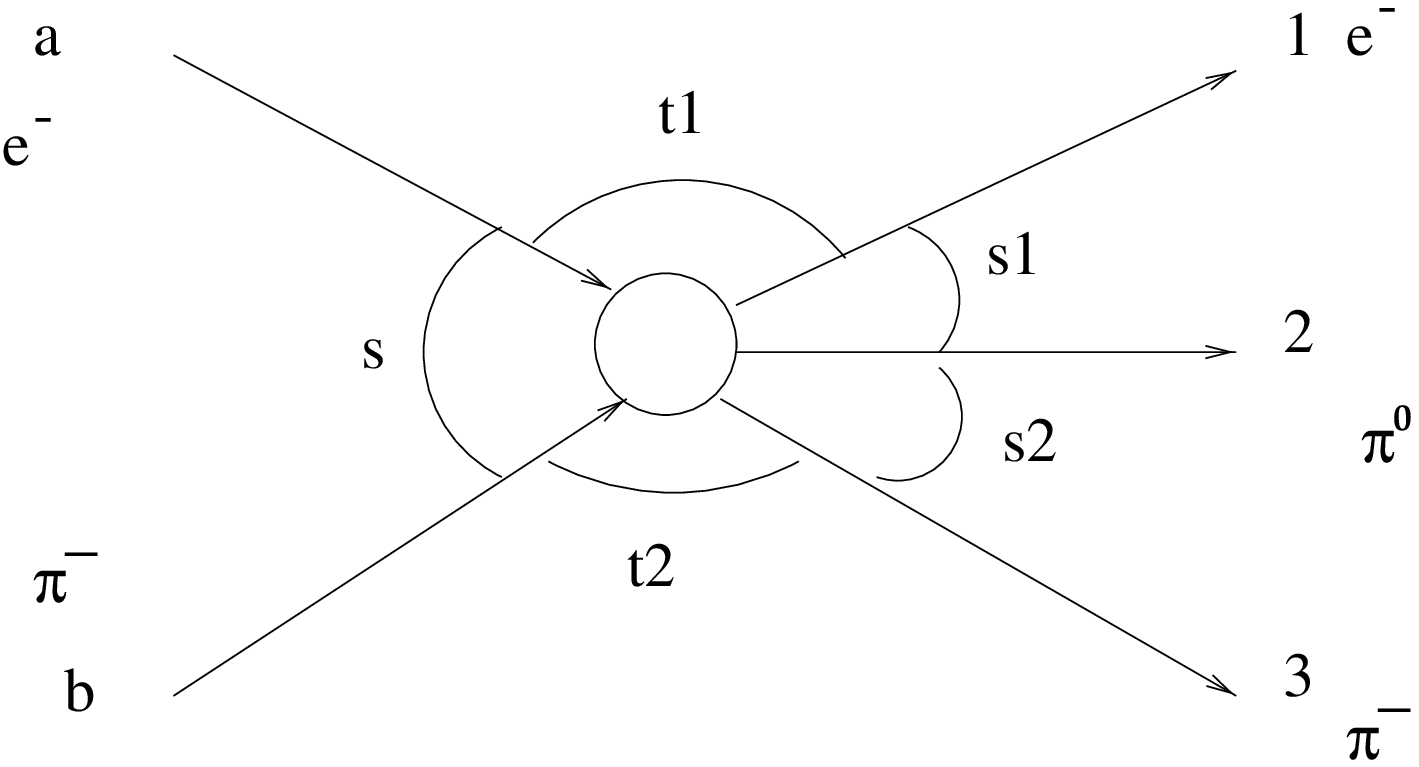,width=10cm}
\caption[]{\label{fig:kinematics}
Kinematics of the reaction
$e^-(p_a)+\pi^-(p_b)\to e^-(p_1)+\pi^0(p_2)+\pi^-(p_3)$.}
\end{figure}

   In this present work,  we will focus on the reaction
\begin{equation}
\pi^- + e^- \rightarrow {\pi^-}'+{e^-}' + \pi^0, \label{eq:pie}
\end{equation}
where an incident high-energy pion scatters inelastically from a target 
electron in an atomic orbit, as shown in Fig.\ \ref{fig:kinematics}.
   This reaction and also $K^-+e^-\to {K^-}'+{e^-}' + \pi^0$
can, in principle, be studied with the (190 - 300) GeV pion and kaon beams in 
the CERN COMPASS experiment \cite{Baum:1996yv}.
   New high-statistics pion data will allow for a determination of the form
factor for $\pi\gamma ^* \rightarrow\pi \pi^{0}$.
   The kaon beam three-flavor process can as well test the expected $N_c$
dependence in a low-energy reaction \cite{Bar:2001qk,Rogalyov:2002bg}.

   A similar (pion) experiment has already been performed at the CERN SPS
\cite{Amendolia:1985bs}.
  The experiment did not explicitly extract a value
${\cal F}_{3\pi}^{\rm exp}$ but rather claimed that the experimental value was
consistent with theory expectations.
   Although the experimental backgrounds were described in
\cite{Amendolia:1985bs}, comparisons of experimental and theoretical
distributions versus different kinematic variables were not shown;
unfortunately, the data are no longer available for such
comparisons \cite{Tenchini}.
   Without presenting such detailed comparisons, Amendolia {\em et al.}
reported 36 events for the reaction 
$\pi^{-} e^-\rightarrow  \pi^{-} e^- \pi^{0}$
corresponding to a cross section of $(2.11 \pm 0.47)$~nb.
   However, without statistical tests such as the Kolmogorov-Smirnov 
distribution test \cite{Press}
comparing experimental and theoretical distributions, it is not possible
to be sure that background events were not included in the cross section
value of Ref.\ \cite{Amendolia:1985bs}.
   The cross section of $\pi^{-}e^-\rightarrow \pi^{-}e^-\pi^{0}$, in 
principle,
may also include $\rho^{-}$ production via the $\pi e \rightarrow \rho e$
transition \cite{LoSecco:1994vt}.
   However, threshold effects eliminate this background for the
300 GeV pion beam energy of \cite{Amendolia:1985bs}, since an energy of
$E_{\pi}\cong 600~\mbox{GeV}$ is required to produce a $\rho$ with
$m_{\rho}=770~\mbox{MeV}$ via $\pi e\rightarrow \rho e$.

   Our work is organized as follows.
   In Sec.\ \ref{section_kinematics} we briefly discuss the kinematics
and formalism of $\pi^{-}e^-\rightarrow \pi^{-}e^-\pi^{0}$.
   In Sec.\ \ref{section_theory} we present the calculation of the
anomaly amplitude ${\cal F}_{3\pi}$ within SU(3) ChPT, discuss some 
phenomenological approaches, and use Monte Carlo
techniques to integrate the cross section and compare it with the experimental
result of \cite{Amendolia:1985bs}.
   In Sec.\ \ref{section_summary} we summarize
our results and draw some conclusions.
   Some technical details are relegated to the Appendices.

\section{\label{section_kinematics} Kinematics and differential cross
section for $\pi^- e^- \rightarrow \pi^- e^- \pi^{0}$}
   Following the nomenclature of Ref.\ \cite{Byckling}, the kinematics for
$a+b\to1+2+3$ is shown in Fig.\ \ref{fig:kinematics} for an incoming pion
that scatters inelastically off an electron target and produces an additional
$\pi^{0}$ in the final state:
$e^-(p_a)+\pi^-(p_b)\to e^-(p_1)+\pi^0(p_2)+\pi^-(p_3)$.
   We consider the target electron to be at rest and the binding energy of
the electrons bound in an atom to be negligible relative to the incoming pion
energy.

   We define the standard set of five invariants as \cite{Byckling}
\begin{eqnarray}
\label{five_invariants}
s_1&\equiv&s_{12}=(p_{1}+p_{2})^{2}=(p_{a}+p_{b}-p_{3})^{2},\nonumber\\
s_{2}&\equiv&s_{23}=(p_{2}+p_{3})^{2}=(p_{a}+p_{b}-p_{1})^{2},\nonumber\\
t_{1}&\equiv&t_{a1}=(p_{a}-p_{1})^{2}=(p_{2}+p_{3}-p_{b})^{2},\nonumber\\
t_{2}&\equiv&t_{b3}=(p_{b}-p_{3})^{2}=(p_{1}+p_{2}-p_{a})^{2},\nonumber\\
s&\equiv&s_{ab}=(p_{a}+p_{b})^{2}=(p_{1}+p_{2}+p_{3})^{2}.
\end{eqnarray}
   The invariant $s$ is fixed by the incident beam energy $E_i$,
$s=M_\pi^2+m_e^2+2 E_i m_e$, and one is left with four scalar variables
$s_1$, $s_2$, $t_1$, and $t_2$.

   The fourfold differential cross section, expressed in terms of the
five invariants of Eq.\ (\ref{five_invariants}), is given by (see,
e.g., Ref.\ \cite{Unkmeir:2001gw})\footnote{Our normalization of
the electron states and of the Dirac spinors is given by
\begin{displaymath}
\langle \vec{p}\,',s'|\vec{p},s\rangle=2
E(\vec{p}\,)(2\pi)^3\delta^3(\vec{p}\,'-\vec{p}\,)\delta_{s's}, \quad
\bar{u}(\vec{p},s')u(\vec{p},s)=2 m_e \delta_{s's}.
\end{displaymath}}
\begin{equation}
\frac{d\sigma}{ds_1ds_2dt_1dt_2}= \frac{\overline{|{\cal
M}|^2}}{4(4\pi)^4 \lambda{(s,m_e^2,M_\pi^2)}(-\Delta_4)^{1/2}},
\label{dsigma}
\end{equation}
where
\begin{equation}
\lambda(x,y,z)=x^2+y^2+z^2-2xy-2xz-2yz, \label{lambda}
\end{equation}
and where the Gram determinant is given by \cite{Byckling}
\begin{equation}
 \Delta_4 =\frac{1}{16}\
  \begin{array}{|cccc|}
  2m_{a}^{2} &2p_{a}\cdot p_{b} &2p_{a}\cdot p_{1} &2p_{a}\cdot p_{3}  \\
  2p_{a}\cdot p_{b} &2m_{b}^{2} &2p_{b}\cdot p_{1} &2p_{b}\cdot p_{3}  \\
  2p_{a}\cdot p_{1} &2p_{b}\cdot p_{1} &2m_{1}^{2} &2p_{1}\cdot p_{3}  \\
  2p_{a}\cdot p_{3} &2p_{b}\cdot p_{3} &2p_{1}\cdot p_{3} &2m_{3}^{2}
  \end{array}\ .
\label{eq:dif19}
\end{equation}
   The factor 1/16 has been extracted for later convenience.
   The expressions of the scalar products entering Eq.\ (\ref{eq:dif19})
in terms of the invariants of Eq.\ (\ref{five_invariants}) are given
in Eq.\ (\ref{scalar_products}).

\begin{figure}
\epsfig{file=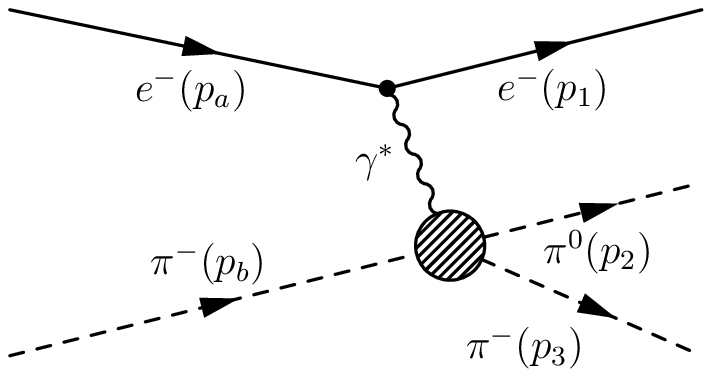,width=8cm}
\caption[]{\label{fig:opea}
One-photon exchange approximation of the reaction
$e^-(p_a)+\pi^-(p_b)\to e^-(p_1)+\pi^0(p_2)+\pi^-(p_3)$.}
\end{figure}

   In the one-photon exchange approximation (see Fig.\ \ref{fig:opea})
the total invariant amplitude ${\cal M}$ can be written as
\begin{equation}
\label{mopea} {\cal M}=-i {\cal F}_{3\pi}(s_2,t_2,u_2;q^2)
\epsilon^\mu F_\mu,
\end{equation}
where $\epsilon^\mu=e\bar{u}(p_1)\gamma^\mu u(p_a)/q^2$ is the virtual
photon polarization vector ($q=p_a-p_1$) and
\begin{equation}
\label{def_Fmu}
F_\mu\equiv \epsilon_{\mu\nu\rho\sigma} p_b^\nu p_2^\rho p_3^\sigma.
\end{equation}
   The squared matrix element of Eq.\ (\ref{dsigma}) involves the
contraction of the standard lepton tensor known from the one-photon
exchange approximation in electroproduction processes,
\begin{equation}
\label{lepton_tensor} \overline{\eta^{\mu\nu}}=\left(2 p_a^\mu
p_1^\nu+2p_1^\mu p_a^\nu+q^2 g^{\mu\nu}\right),
\end{equation}
with the hadronic tensor and is given by
\begin{equation}
\label{M2overline} \overline{|{\cal M}|^2}=
\left(\frac{e}{q^2}\right)^2 |{\cal F}_{3\pi}|^2\,
\overline{\eta^{\mu\nu}} F_\mu F_\nu =
\left(\frac{e}{q^2}\right)^2 |{\cal F}_{3\pi}|^2 \left(4 p_a\cdot
F\, p_1\cdot F +q^2 F\cdot F\right).
\end{equation}
   The explicit expression for Eq.\ (\ref{M2overline}) is given in
App.\ \ref{squared_amplitude}.

\section{\label{section_theory}
Theoretical description of the ${\cal F}_{3\pi}$ amplitude}
   In this section we describe the theoretical input to
our analysis of the reaction $\pi^{-}e^-\rightarrow \pi^{-}e^-\pi^{0}$.
   Since we work in the one-photon exchange approximation, it is sufficient
to consider the transition-current matrix element
\begin{displaymath}
\langle \pi^0(p_2),\pi^-(p_3)|J_\mu(0)|\pi^-(p_b)\rangle ={\cal
F}_{3\pi}(s_2,t_2,u_2;q^2)\epsilon_{\mu\nu\rho\sigma} p_b^\nu
p_2^\rho p_3^\sigma,
\end{displaymath}
where $J_\mu$ is the electromagnetic current operator (including the
elementary charge).
   In the isospin-symmetric limit, ${\cal F}_{3\pi}$ is a completely
symmetric function of the Mandelstam variables $s_2$, $t_2$, and
$u_2$.
   In the {\em physical} region, the Mandelstam variables satisfy the
standard relation $s_2+u_2+t_2=2 M_{\pi^-}^2+M_{\pi^0}^2+q^2$.
   We will lay emphasis on a calculation within the framework of 
chiral perturbation theory at ${\cal O}(p^6)$ but will also discuss
the results of some (more) phenomenological approaches.
   This will allow us to have an estimate of effects which would 
be subsumed in terms of ${\cal O}(p^8)$ and higher.

\subsection{Chiral perturbation theory at ${\cal O}(p^6)$} 
\label{sub_sec_chpt}
\begin{figure}
\epsfig{file=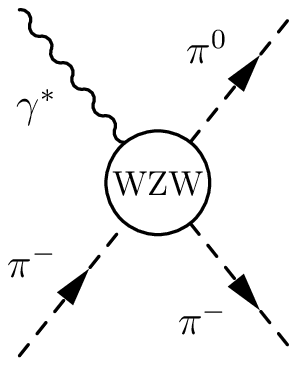,width=4cm}
\caption[]{\label{fig:tree_4}
WZW diagram obtained from Eq.\ (\ref{L3phigamma_WZW}).}
\end{figure}

   Besides the neutral-pion decay into two photons, the amplitude
for $\gamma+\pi^-\to \pi^0+\pi^-$ is of  
prime interest for testing our understanding of anomalous Ward identities.
   In the momentum and quark-mass expansion, its leading-order contribution
is of ${\cal O}(p^4)$ and originates from the Wess-Zumino-Witten
action \cite{Wess:yu,Witten:tw}.
   The interaction Lagrangian relevant in the presence of external 
electromagnetic fields (described by the vector potential ${\cal A}_\mu$) 
is given by \cite{Wess:yu,Witten:tw,Scherer:2002tk}
\begin{equation}
\label{lanoelm}
{\cal L}^{\rm e.m.}_{\rm WZW}=-e {\cal A}_\mu J^\mu
+i \frac{ e^2 }{16\pi^2}\epsilon^{\mu\nu\rho\sigma}
\partial_\nu {\cal A}_\rho{\cal A}_\sigma
\mbox{Tr}[2Q^2(U\partial_\mu U^\dagger - U^\dagger \partial_\mu U)
- Q U^\dagger Q \partial_\mu U 
+Q U Q \partial_\mu U^\dagger],
\end{equation}
where $Q=\mbox{diag}(2/3,-1/3,-1/3)$ denotes the quark-charge matrix
and 
\begin{equation}
\label{matrixU}
U=\exp\left(i\frac{\phi}{F_0}\right),\quad \phi=\sum_{a=1}^8 \lambda_a \phi_a,
\end{equation}
contains the Goldstone boson fields.
   The current
\begin{equation}
\label{jmu}
J^\mu=\frac{\epsilon^{\mu\nu\rho\sigma}}{16\pi^2}
\mbox{Tr}(Q\partial_\nu U U^\dagger \partial_\rho U U^\dagger
\partial_\sigma U U^\dagger
+Q U^\dagger \partial_\nu U U^\dagger \partial_\rho U U^\dagger
\partial_\sigma U),
\end{equation}
by itself is not gauge invariant and the additional terms of 
Eq.\ (\ref{lanoelm}) are required to obtain a gauge-invariant action.
   The first term of Eq.\ (\ref{lanoelm}) gives rise to the $3\phi+\gamma$
coupling (see Fig.~\ref{fig:tree_4})
\begin{equation}
\label{L3phigamma_WZW}
{\cal L}^{3\phi+\gamma}_{\rm WZW}
=ie\frac{\epsilon^{\mu\nu\rho\sigma}}{8\pi^2 F_0^3}
{\cal A}_\mu\mbox{Tr}(Q\partial_\nu\phi\partial_\rho\phi\partial_\sigma\phi),
\end{equation}   
whereas the second is responsible for the $\pi^0\to\gamma\gamma$ decay
not discussed in this paper.

   Weinberg's power counting scheme \cite{chpt} establishes a connection 
between the chiral expansion and the loop expansion. 
   Since the anomalous sector only starts at ${\cal O}(p^4)$, 
the contribution at ${\cal O}(p^6)$ results from either one-loop diagrams 
with exactly one vertex from the WZW term or tree-level diagrams with 
exactly one vertex from the anomalous Lagrangian at ${\cal O}(p^6)$.
   In order to determine the one-loop contributions we need, besides
Eq.\ (\ref{lanoelm}), the WZW contribution involving 5 Goldstone
bosons,
\begin{equation}
\label{lwzw5phi}
{\cal L}_{\rm WZW}^{5\phi}=
\frac{1}{80\pi^2 F^5_0}
\epsilon^{\mu\nu\rho\sigma}\mbox{Tr}(\phi\partial_\mu\phi\partial_\nu\phi
\partial_\rho\phi\partial_\sigma\phi),
\end{equation}
and the lowest-order Lagrangian in the presence of
an electromagnetic field,
\begin{equation}
\label{lowlag}
{\cal L}_2=\frac{F_0^2}{4}\mbox{Tr}[D_\mu U (D^\mu U)^\dagger]
+\frac{F^2_0}{4}\mbox{Tr}(\chi U^\dagger + U\chi^\dagger),
\end{equation}
where the relevant covariant derivative is given by
$D_\mu U=\partial_\mu U +ie {\cal A}_\mu [Q,U]$ and 
$\chi=2 B_0 M$ contains the quark-mass matrix $M$ and $B_0$ is related
to the scalar quark condensate in the chiral limit.
   The most general anomalous Lagrangian at ${\cal O}(p^6)$ has recently
been derived in Refs.\ \cite{Ebertshauser:2001nj,Bijnens:2001bb}. 
   According to Table V of Ref.\ \cite{Ebertshauser:2001nj}, seven
structures have the potential of contributing to $3\phi+\gamma$ vertices.
   In principle, the corresponding low-energy coupling constants should
be calculable from the underlying theory. 
   However, since we cannot yet solve QCD, the parameters are either
taken as free parameters that are fitted to experimental data or are
estimated from models such as meson-resonance saturation
\cite{Ecker:yg,Ecker:1988te}.
    
   In what follows, we will make use of the SU(3) version of chiral 
perturbation theory, because this will allow us in future calculations
to make contact with other anomalous processes involving in addition kaons.
   Moreover, we note that previous calculations at ${\cal O}(p^6)$ 
\cite{Bijnens:1989ff} were performed for real photons, $q^2=0$, because
the amplitude was embedded in a  Primakoff reaction, where the virtual 
photon of the Coulomb field of a heavy nucleus is quasi real.
   For our reaction such an approximation is not admissable which we 
will also explicitly see when we discuss the results.

\begin{figure}
\epsfig{file=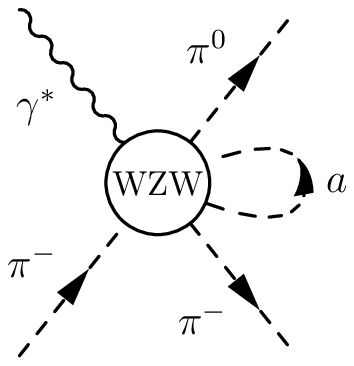,width=4cm}
\caption[]{\label{fig:loop_a}
One-loop diagram obtained from expanding the first term of
Eq.\ (\ref{lanoelm}) to fifth order in the Goldstone boson fields
and contracting two lines to form a loop.}
\end{figure}

\begin{figure}
\epsfig{file=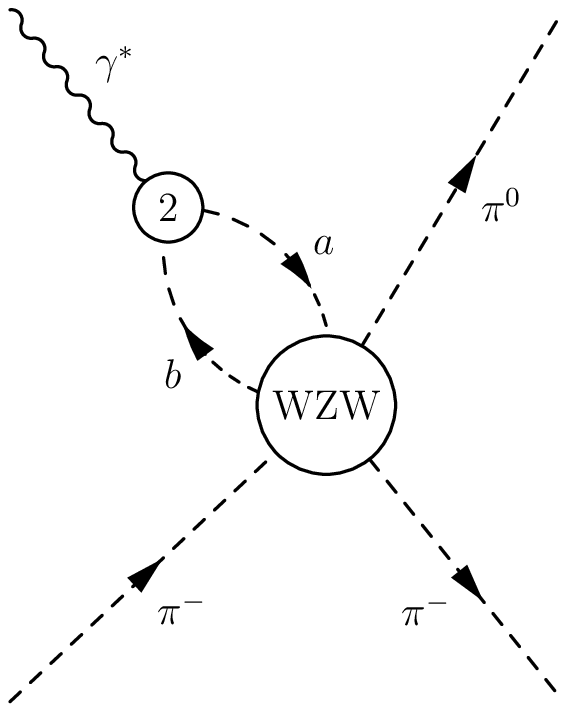,width=4cm}
\caption[]{\label{fig:loop_b}
One-loop diagram obtained from contracting two lines of Eq.\ (\ref{lwzw5phi})
with the two lines of the $2\phi+\gamma$ vertex from
${\cal L}_2$ to form a loop.}
\end{figure}

\begin{figure}
\epsfig{file=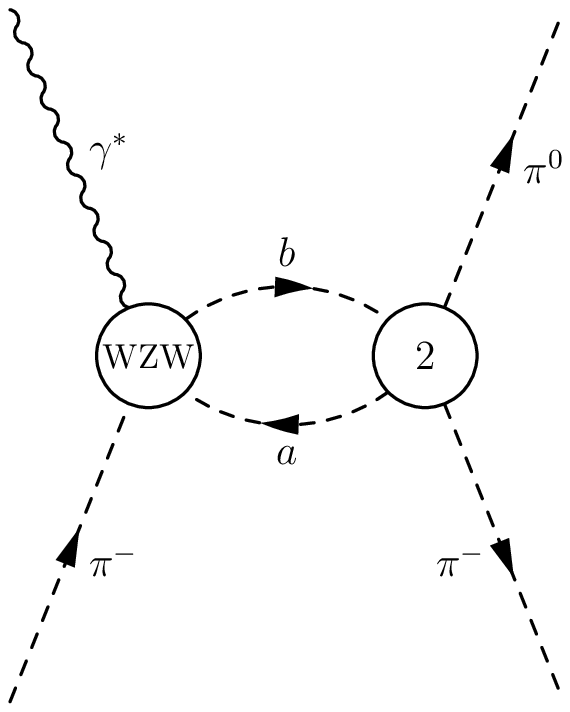,width=4cm}\hspace{2em}
\epsfig{file=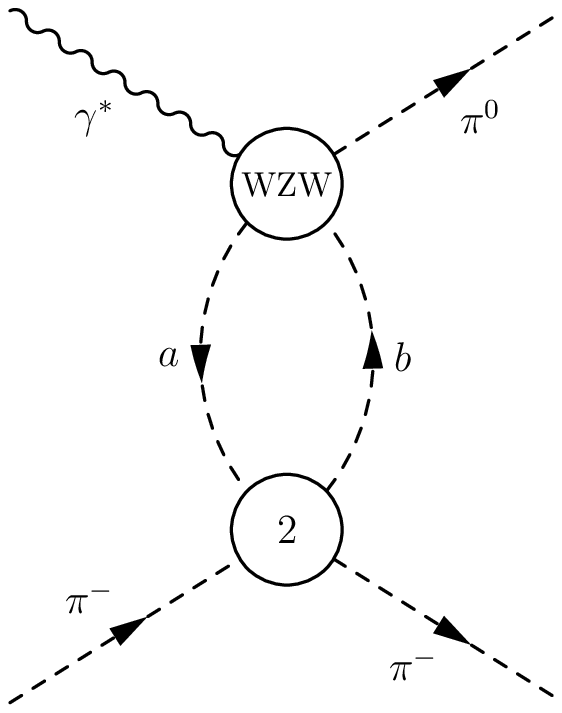,width=4cm}\hspace{2em}
\epsfig{file=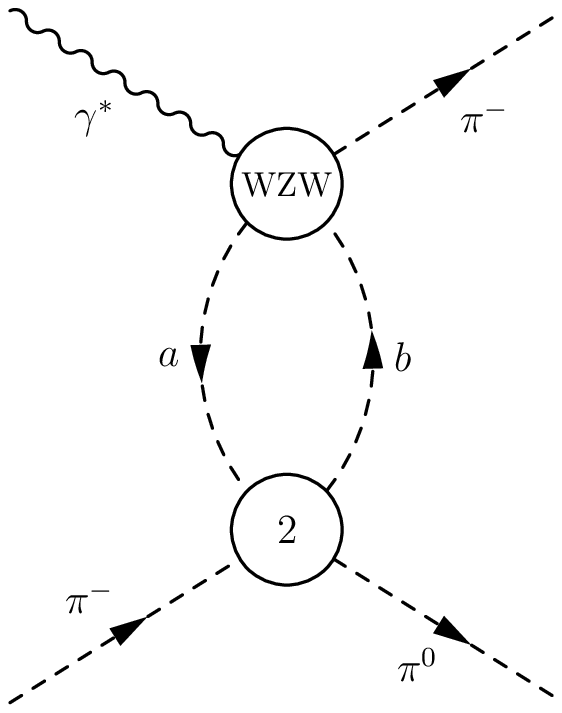,width=4cm}
\caption[]{\label{fig:loop_cde}
One-loop diagrams obtained from contracting two lines
of Eq.\ (\ref{L3phigamma_WZW}) with two lines of 
the $4\phi$ vertex from ${\cal L}_2$ to form a loop (cuts in the
$s_2$, $t_2$, and $u_2$ channels, respectively).}
\end{figure}

\begin{figure}
\epsfig{file=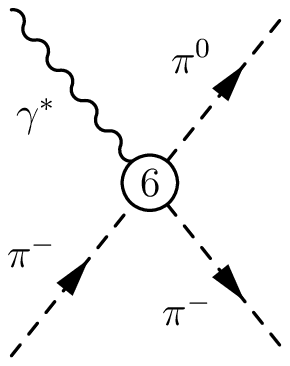,width=4cm}
\caption[]{\label{fig:tree_6}
Contact diagram obtained from ${\cal L}_6$.}
\end{figure}

   The relevant one-loop diagrams are shown in Figs.~\ref{fig:loop_a},
\ref{fig:loop_b}, and \ref{fig:loop_cde} and fall into
two distinct groups. 
   The first category just includes one graph (Fig.~\ref{fig:loop_a})
whose loop is attached to one single vertex (the loop is composed by
one internal line so to speak), while in the second category the loop always
binds two different vertices together (the loop is therefore composed
of two internal lines).
   Moreover, at ${\cal O}(p^6)$ one obtains a contact contribution 
shown in Fig.~\ref{fig:tree_6}.
   Combining the ${\cal O}(p^4)$ and ${\cal O}(p^6)$ results,
multiplying with a factor of $\sqrt{Z_\pi}$ for each external pion line,
and renormalizing the coefficients of the ${\cal O}(p^6)$ Lagrangian, 
the result of the one-loop calculation in SU(3) ChPT at
${\cal O}(p^6)$ is given by
\cite{Ebertshaeuser:2001}:
\begin{eqnarray}
{\cal F}_{3\pi}(s_2,t_2,u_2;q^2)&=&\frac{e}{4 \pi^{2}F^{3}_\pi}\left(1 +
C_{M_\pi^2} M_\pi^2 + C_{q^2}q^2
+\frac{1}{32\pi^{2}F_{\pi}^{2}}\left\{
\frac{s_2+u_2+t_2}{3}\ln\left(\frac{\mu^2}{M_{\pi}^{2}}\right)
\right.\right.\nonumber\\
&& +q^2\ln\left(\frac{\mu^{2}}{M_K^{2}}\right)
+\frac{5}{9}(s_2+u_2+t_2+3q^2)\nonumber\\
&&\left.\left. +\frac{4}{3}\left[F(s_2,M_\pi^2)+F(t_2,M_\pi^2)
+F(u_{2},M_\pi^2)+3F(q^2,M_K^2)\right]\right\}\right). \label{eq:theor3}
\end{eqnarray}
   The constants $C_{M_\pi^2}$ and $C_{q^2}$ are linear combinations
of {\em renormalized} low-energy coupling constants $\hat{L}^{6,\epsilon}_{i}$
of the most general
odd-intrinsic-parity Lagrangian at ${\cal O}(p^6)$
\cite{anop6,Ebertshauser:2001nj,Bijnens:2001bb,Ebertshaeuser:2001},
\begin{eqnarray}
\label{cdef} C_{M_\pi^2}&=&512
\pi^2(\hat{L}^{6,\epsilon}_{13}-\hat{L}^{6,\epsilon}_{14}
-2\hat{L}^{6,\epsilon}_{5}-\hat{L}^{6,\epsilon}_{6}),\nonumber\\
C_{q^2}&=&-\frac{512
\pi^2}{3}(\hat{L}^{6,\epsilon}_{13}-\hat{L}^{6,\epsilon}_{14}).
\end{eqnarray}
   These coefficients still depend on the renormalization scale $\mu$ but in
such a way that the scale dependence of the logarithms in 
Eq.\ (\ref{eq:theor3}) is precisely compensated.
   The function $F$ originates from a standard one-loop integral of
mesonic chiral perturbation theory and is given by
\begin{eqnarray}
F(a,m^2)&\equiv&m^2\left(1-\frac{a}{4m^2}\right)
J^{(0)}\left(\frac{a}{m^2}\right)
-\frac{a}{2}\ ,\nonumber\\
\label{fktJ0}
J^{(0)}(x)
&\equiv& \int_0^1 dz\ln[1+x(z^2-z)-i0^+]\nonumber\\
&=&\left \{ \begin{array}{ll}
-2-\sigma\ln\left(\frac{\sigma-1}{\sigma+1}\right) & (x<0) \nonumber\\
-2+2\sqrt{\frac{4}{x}-1}\,\mbox{arccot}
\left(\sqrt{\frac{4}{x}-1}\right) &  (0\le x<4) \nonumber\\
-2-\sigma\ln\left(\frac{1-\sigma}{1+\sigma}\right)-i\pi\sigma
& (x>4)
\end{array}, \right.\nonumber
\end{eqnarray}
with $\sigma(x)\equiv\sqrt{1-4/x}$ for $x\notin [0,4]$.

   Up until now, we have carried out everything which is necessary to
meet the requirements of a consistent ${\cal O}(p^6)$ calculation within
the framework of mesonic ChPT. 
   However, our result still depends on two unknown parameters 
($C_{M_\pi^2}$ and $C_{q^2}$) which prevent us from {\em predicting}
observables such as the total cross section or distributions.
   Of course, these low-energy coupling constants (LECs) can in principle be 
determined within appropriate experiments.
   Here, we actually move on and estimate the constants by using
theoretical means which evidently have to go beyond mesonic ChPT.
   The LECs are supposed to include whatever QCD information on all particles
which do not belong to the Goldstone boson octet.
   At low energies the lightest are expected to be significant 
and we are thus led to consider the effects due to the vector mesons
\cite{Ecker:yg,Ecker:1988te}.
   For that purpose we made use of the nonlinear chiral 
Lagrangian of Ref.\ \cite{Fujiwara:1984mp}, evaluated the tree-level
diagrams contributing to $\gamma^\ast+\pi^-\to \pi^0+\pi^-$  involving
internal vector meson lines, and expanded the propagators to be able
to collect the arising ${\cal O}(p^6)$ pieces (for later purposes we
also consider the expressions keeping the full propagators).
   A comparison (matching) with the polynomial $p^6$ pieces of the
most general anomalous Lagrangian yields the estimate
\begin{equation}
\hat{L}^{6,\epsilon}_{5}=-\frac{3}{1024 \pi^2 m_V^2}=\hat{L}^{6,\epsilon}_{13},
\quad
\hat{L}^{6,\epsilon}_{6}=-\hat{L}^{6,\epsilon}_{14}=3\hat{L}^{6,\epsilon}_{5}.
\end{equation}
   Using $m_V=m_\rho=m_\omega$ in SU(3) results in
\begin{equation}
C_{M_\pi^2}=\frac{3}{2m_\rho^2},\quad C_{q^2}=\frac{2}{m_\rho^2}. 
\label{cestimate}
\end{equation}
   Instead of expanding the vector-meson propagators in the vector-meson 
saturation calculation, we could also keep the complete propagators.
   This would correspond to the replacement
\begin{equation}
\label{replacement} C_{M_\pi^2} M_\pi^2 + C_{q^2}q^2= \frac{3
M_\pi^2+4q^{2}}{2m_{\rho}^{2}} \to \frac{1}{2}
\left(\frac{s_2}{m_{\rho}^{2}-s_2}+
\frac{t_2}{m_{\rho}^{2}-t_2}+\frac{u_2}{m_{\rho}^{2}-u_2}\right) +\frac{3}{2}
\frac{q^2}{m_\omega^2-q^2}
\end{equation}
in Eq.\ (\ref{eq:theor3}).
   We then obtain some estimate of higher-order terms beyond ${\cal O}(p^6)$.

\begin{figure}
\epsfig{file=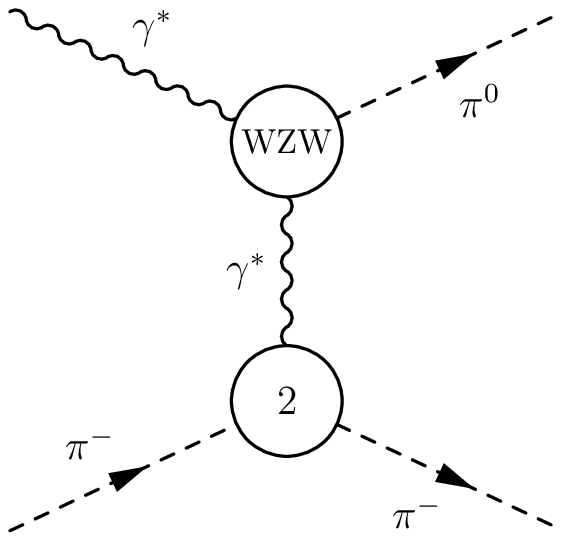,width=6cm}
\caption[]{\label{fig:em_corr}
Electromagnetic correction.}
\end{figure}

   A full calculation of all ${\cal O}(e^2)$ radiative corrections as
well as the isospin symmetry breaking effects due to the different $u$- and 
$d$-quark masses is beyond the scope of the present paper.
   As discussed in Ref.\ \cite{Ametller:2001yk}, the most important
electromagnetic correction originates from a photon-photon fusion into a 
neutral pion (see Fig.\ \ref{fig:em_corr}) yielding an additional 
contribution of the type
\begin{equation}\label{deltae2}
\Delta {\cal F}_{3\pi}^{(e^2)}=\frac{e}{4\pi^2 F_\pi^3}\left(-\frac{2e^2
F_\pi^2}{t_2}\right).
\end{equation}
   Due to the $1/t_2$ pole of the exchanged photon, Eq.\ (\ref{deltae2})
becomes important for small values of $t_2$.
   Note that including the single diagram of Fig.\ \ref{fig:em_corr} does
not lead to a conflict with gauge invariance.
   On the other hand, other electromagnetic corrections were found to be very 
small in Ref.\ \cite{Ametller:2001yk}, where the authors concluded that
their full calculation can be very well reproduced by adding only
the contribution of Eq.\ (\ref{deltae2}).

\subsection{Phenomenological approaches}
   In our analysis we will also compare data with phenomenological calculations
using three different forms of the transition-current matrix element:
\begin{enumerate}
\item Phenomenological ansatz of Terent'ev \cite{Terent'ev:1971kt}:
\begin{equation}
{\cal F}_{3\pi}(s_2,t_2,u_2;q^2)={\cal
F}_{3\pi}^{(0)}\left[1+\Delta_{\rho}
\left(\frac{s_2}{m_{\rho}^{2}-s_2}+
\frac{t_2}{m_{\rho}^{2}-t_2}+\frac{u_2}{m_{\rho}^{2}-u_2}\right)
+\Delta_{\omega}\frac{q^2}{m_{\omega}^{2}-q^2} \right].
\label{eq:theor1}
\end{equation}
   Here, the multiplicative constant ${\cal F}_{3\pi}^{(0)}$
refers to the low-energy prediction of Eq.\ (\ref{eq:int5}) and the variation 
of the function ${\cal F}_{3\pi}$ is supposed to come mainly from
vector-meson-exchange diagrams.
   The parameters $\Delta_\rho$ ($\Delta_\omega$) implicitly
contain factors of $1/{\cal F}_{3\pi}^{(0)}$ and are related to the partial
widths $\Gamma(\rho^+\to\pi^+\pi^0)$ and $\Gamma(\rho\to\pi\gamma)$
[$\Gamma(\omega\to e^+ e^-)$ and $\Gamma(\omega\to3\pi)$].
   In the numerical analysis we make use of $m_\rho=770$ MeV and
$m_\omega=782$ MeV.
   Comparing Eq.\ (\ref{eq:theor1}) with Eq.\ (\ref{replacement}) 
would result in $\Delta_\rho\approx 1/2$ and
$\Delta_\omega\approx 3/2$ in comparison with $\Delta_\rho\lesssim 1/2$ and
$\Delta_\omega\approx 2.6$ of Ref.\ \cite{Terent'ev:1971kt}.

\item Pole model of Rudaz including vector-meson dominance (VMD)
\cite{Amendolia:1985bs}:
\begin{equation}
{\cal F}_{3\pi}(s_2,t_2,u_2;q^2)={\cal
F}_{3\pi}^{(0)}\frac{m_\omega^2}{m_{\omega}^{2}-q^2}\frac{m_\rho^2}{3}
\left[\frac{1}{m_{\rho}^{2}-s_2}+
\frac{1}{m_{\rho}^{2}-t_2}+\frac{1}{m_{\rho}^{2}-u_2}\right].
\label{eq:theorudaz}
\end{equation}

\item VMD model including the effects of final state p-wave
$\pi\pi$ scattering \cite{Holstein:1995qj}:
\begin{eqnarray} {\cal F}_{3\pi}(s_2,t_{2},u_2) &=&
-\frac{1}{2}\frac{e}{4 \pi^{2}F^{3}_\pi}
\left[1-\left(\frac{m_{\rho}^{2}}{m_{\rho}^{2}-s_2}+
\frac{m_{\rho}^{2}}{m_{\rho}^{2}-t_{2}}+\frac{m_{\rho}^{2}}
{m_{\rho}^{2}-u_2}\right)\right]\nonumber \\
&&\times
\left(\frac{1-s_2/m_{\rho}^{2}}{D_{1}(s_2)}\right)
\left(\frac{1-t_{2}/m_{\rho}^{2}}{D_{1}(t_{2})}\right)
\left(\frac{1-u_2/m_{\rho}^{2}}{D_{1}(u_2)}\right) ,
\label{eq:theor5}
\end{eqnarray}
where
\begin{equation}
D_{1}(a)=1-\frac{a}{m_{\rho}^{2}}-\frac{a}
{96\pi^{2}F_{\pi}^{2}}\ln\left(\frac{m_{\rho}^2}
{M_{\pi}^{2}}\right)-\frac{1}{24\pi^{2}F^{2}_{\pi}}F(a,M_\pi^2).
\label{eq:theor6}
\end{equation}
Note that the ansatz of Eq.\ (\ref{eq:theor5}) has only been
derived for real photons, $q^2=0$, and we therefore have to 
expect some shortcomings in the description of 
$\gamma^\ast+\pi^-\to\pi^0+\pi^-$.
\end{enumerate}

\subsection{Results and discussion}

   Using Monte Carlo techniques we determined the total cross section based 
on Eq.\
(\ref{dsigma}) for kinematical variables inside the region specified by
\begin{eqnarray}
&&0.0184\, \mbox{GeV}^2<s_1<0.186\,\mbox{GeV}^2,\\
&&0.0754\, \mbox{GeV}^2<s_2<0.325\,\mbox{GeV}^2,\\
&&-0.236\, \mbox{GeV}^2<t_1<-0.001\,\mbox{GeV}^2=t_1^{\rm cut},\\
&&-0.269\, \mbox{GeV}^2<t_2<0, \label{kinreg}
\end{eqnarray}
which are the minimal (maximal) values obtained from the equations for the
kinematical boundaries [see Ref.\ \cite{Byckling} and Eq.\ (\ref{kinbound})].
   For the generated invariants the positivity of $-\Delta_4$ of 
Eq.\ (\ref{dsigma})
is checked and events with positive $\Delta_4$ are rejected.
   In order to check the Monte Carlo calculations we also compared the result 
with an explicit numerical integration using the simplifications of a 
constant ${\cal F}_{3\pi}$ and $m_e^2\to 0$ 
(see Appendix \ref{appendix_direct_calculation}).

   The results for the total cross section are shown in Table
   \ref{table_sigmatot}.
   The first column denotes the model/theory and the corresponding parameters 
used; the second column contains the integrated cross section for each case 
with ${\cal F}_{3\pi}^{(0)}$ fixed to 
$e/(4\pi^{2}F^{3}_\pi)=9.72\,\mbox{GeV}^{-3}$.
   The cases 1), 4), 5), and 6) of Table \ref{table_sigmatot} were already used
in Fig.~4 of Ref.\ \cite{Amendolia:1985bs} and our corresponding cross sections
are in reasonably good agreement.
   In the third column we show the respective physical threshold amplitudes.
   In the chiral limit, the threshold amplitudes should reduce to the 
low-energy prediction $e/(4\pi^{2}F^{3}_0)$ of Eq.\ (\ref{eq:int5}).
   In this context it is important to realize that, in general, the dependence
of the threshold amplitude on $M_\pi^2$ results from both kinematical variables
and explicit symmetry breaking.
   The models of 2) - 6) can only account for the first type of dependence
and the corresponding modification is of the type 
$const.\times M_\pi^2/m_\rho^2$, where the relevant constant depends on the 
parameters of the model and is of the order of 1.
   The results of 7) - 11), in addition, contain corrections from Goldstone
boson loops of the form $const.'\times M_\pi^2/(4\pi F_\pi)^2$ which are of 
both kinematical and chiral symmetry breaking type. 
   In the calculation of 9), the $1/t_2$ singularity due to the 
electromagnetic correction of Eq.\ (\ref{deltae2}) generates a 17 \% increase 
in units of $e/(4\pi^{2}F^{3}_\pi)$,
but when integrated over $t_2$ ultimately leads to a less pronounced 
contribution to the cross section.
   Finally, in the last column we have also included the overall factor
${\cal F}_{3\pi}^{(0)\rm extr}$ which one extracts based on the experimental
result $(2.11 \pm 0.47)$~nb of Ref.\ \cite{Amendolia:1985bs} if one treats 
${\cal F}_{3\pi}^{(0)}$ as a free parameter in the respective model. 
The error in ${\cal F}_{3\pi}^{(0)\rm extr}$ only reflects the error in the 
experimental cross section and does not include any error estimate implied by 
the models. 
   However, we explicitly do {\em not} advocate such an extraction as a strict
test of the low-energy theorem of Eq.\ (\ref{eq:int5}), because it introduces 
a bias in how the chiral limit is approached.
   Rather, at this point, the only rigorous approach consists of
using the chiral expansion as in Eq.\ (\ref{eq:theor3}) and confronting it
directly with experimental results.

   Using the estimate of Eq.\ (\ref{cestimate}) for the parameters
$C_{M_\pi^2}$ and $C_{q^2}$, we obtain as the ChPT
result at ${\cal O}(p^6)$
\begin{equation}
\sigma=2.05\,\mbox{nb}.
\end{equation}
   By also including the most prominent electromagnetic correction 
\cite{Ametller:2001yk} in terms of photon-photon fusion into a neutral pion 
the result increases slightly:
\begin{equation}
\sigma=2.17\,\mbox{nb}.
\end{equation}
   Both results are in excellent agreement with the experimental result
$(2.11 \pm 0.47)$~nb of Ref.\ \cite{Amendolia:1985bs}.
   In general, the conceptual advantage of the ChPT calculation over the 
remaining empirical models is that it is the only calculation which naturally 
incorporates both genuine quantum effects (loops) and higher-order 
corrections (as estimated from the VMD saturation).
   Moreover, in principle, a controlled improvement is possible by performing a
complete ${\cal O}(p^8)$ calculation, whereas the remaining calculations suffer
from the absence of a systematic method of improvement.
   A comparison of 8) with 7) clearly shows the
necessity to include the consequences resulting from the virtuality of the
exchanged photon.
   This can also be seen from the transitions 2) to 3) or 4) to 5) in the
calculations using Terent'ev's model.
   On the other hand, the spread of the obtained cross sections is an 
indication that higher-order terms (in the chiral expansion) may play an 
important role.
   This conjecture is supported by an analysis for a pion beam energy of 150 
GeV which leads to total cross section results between 0.17 nb and 0.21 nb, 
i.e., the results scatter substantially less than for the higher energy.

\begin{figure}
\epsfig{file=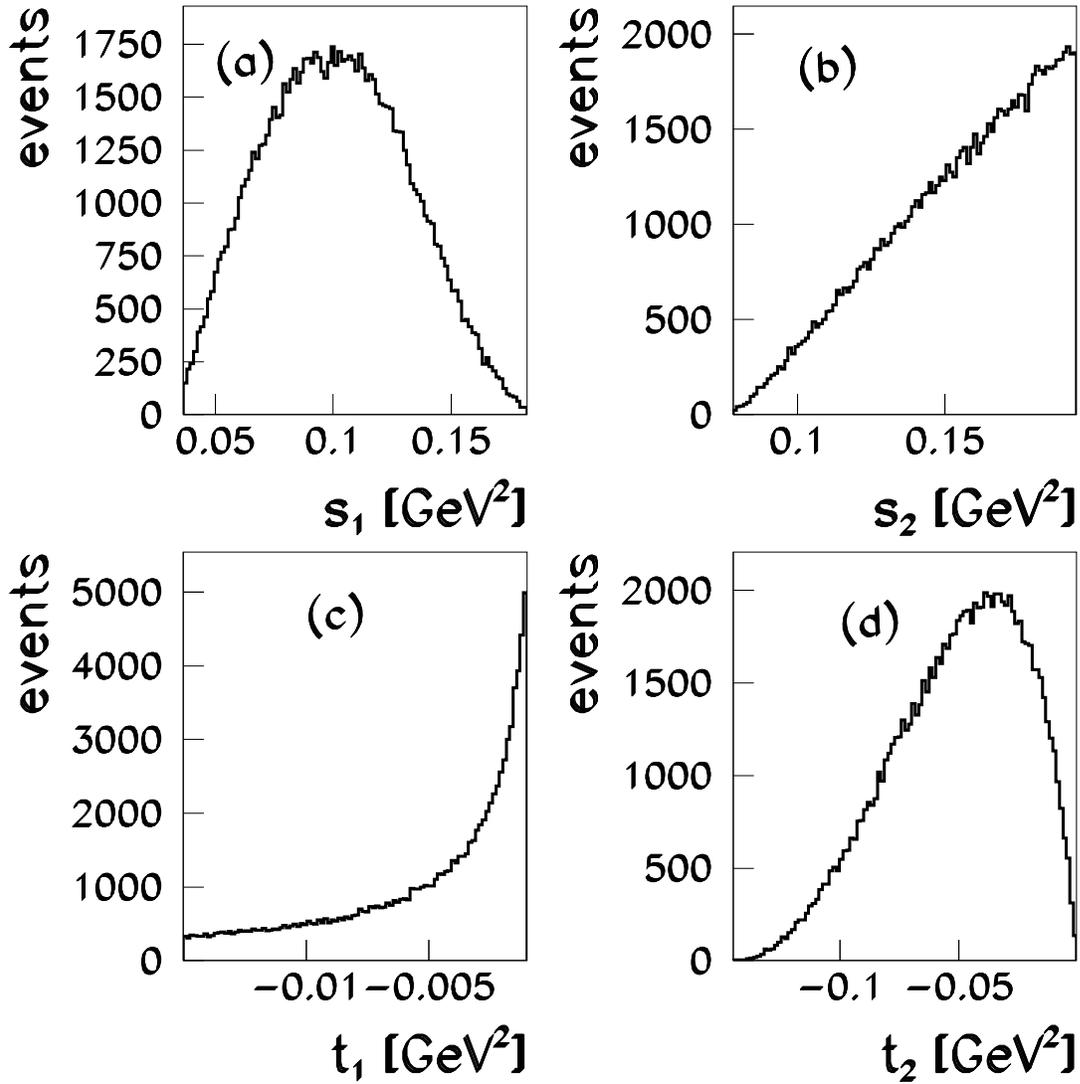,width=\textwidth}
\caption[]{\label{fig:invs}
   The distributions of events as functions of the invariants 
(a) $s_1$, (b) $s_2$, (c) $t_1$, and (d) $t_2$ as obtained using chiral 
perturbation theory at ${\cal O}(p^6)$ [see Eq.\ (\ref{eq:theor3})] with 
the low-energy constants of Eq.\ (\ref{cestimate}) and including the most 
prominent electromagnetic correction of Eq.\ (\ref{deltae2}).}
\end{figure}

\begin{figure}
\epsfig{file=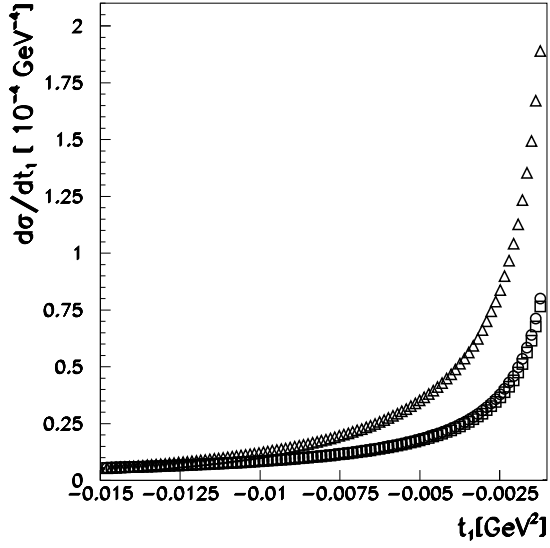,width=0.5\textwidth}
\caption[]{\label{fig:mv_dif} Differential cross section
$d\sigma{/}dt_1$ as a function of $t_1$ using chiral perturbation theory
at ${\cal O}(p^6)$ [see Eq.\ (\ref{eq:theor3})] including
the most prominent electromagnetic correction of
Eq.\ (\ref{deltae2}).
   The low-energy constants of Eq.\ (\ref{cestimate}) have been 
fixed using $M_V=0.2$ GeV (triangles), $M_V=m_{\rho}$ (circles), and
$M_V=2 m_{\rho}$ (squares), respectively.} 
\end{figure}

\begin{figure}
\epsfig{file=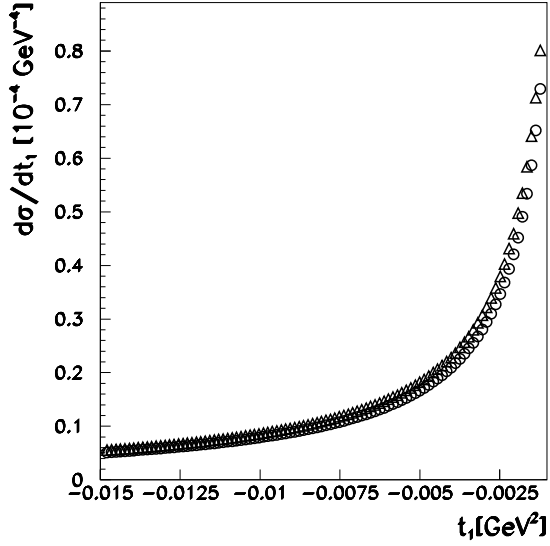,width=0.5\textwidth}
\caption[]{\label{fig:mod_8_9}
Differential cross section $d\sigma{/}dt_1$ as a function of $t_1$ 
using chiral perturbation theory at ${\cal O}(p^6)$ 
[see Eq.\ (\ref{eq:theor3})] with the low-energy constants of 
Eq.\ (\ref{cestimate}) without (circles) and including (triangles)
the most prominent electromagnetic correction of
Eq.\ (\ref{deltae2}).}
\end{figure}

\begin{figure}
\epsfig{file=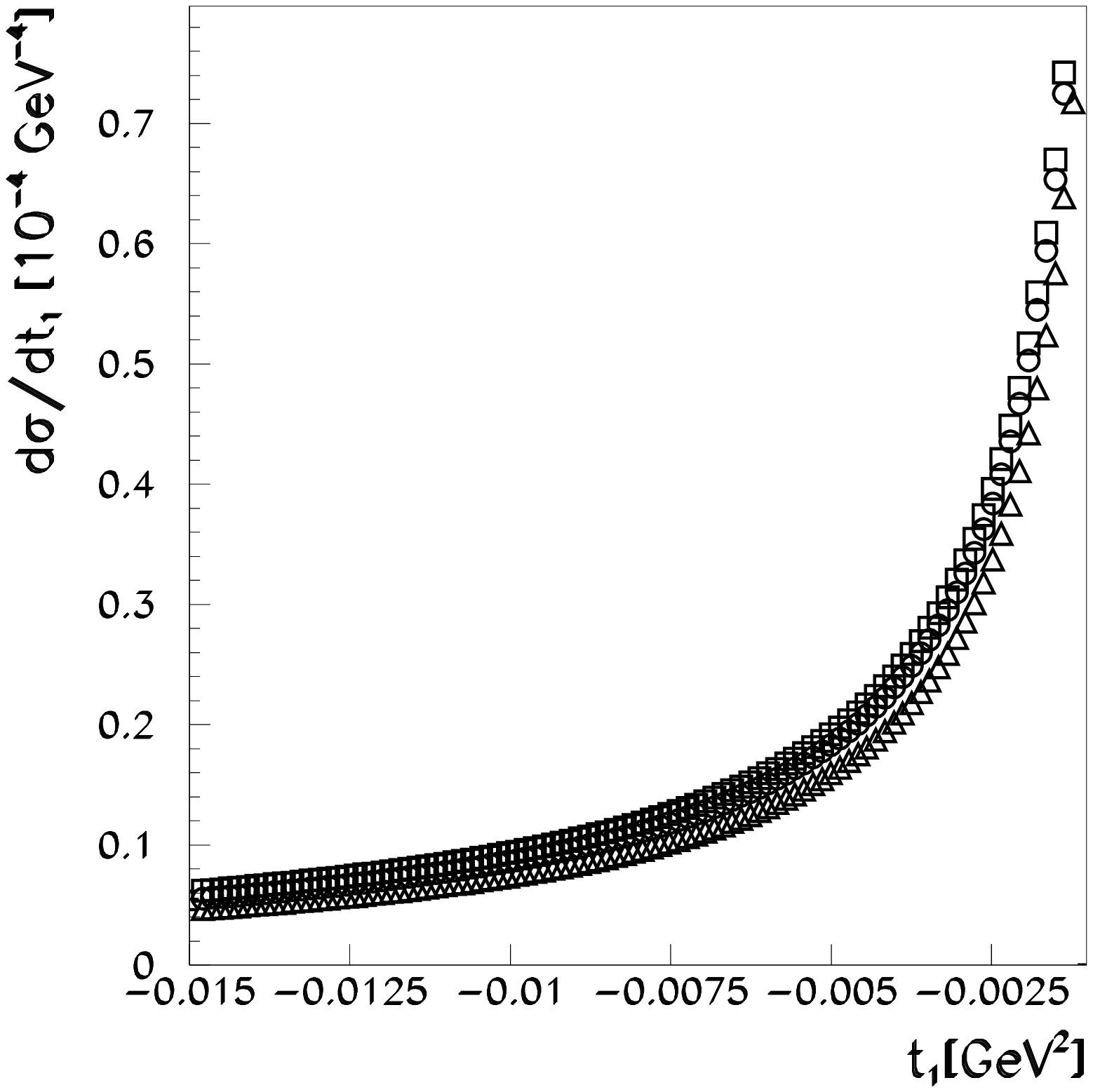,width=0.5\textwidth}
\caption[]{\label{fig:mod_comp}
Differential cross section $d\sigma{/}dt_1$ as a function of $t_1$ for 
the model of Terent'ev, Eq.\ (\ref{eq:theor1}) with $\Delta_\rho=0.35$ and
$\Delta_\omega=3.2$ (triangles), 
ChPT at ${\cal O}(p^6)$ [Eq.\ (\ref{eq:theor3})] 
plus electromagnetic correction of Eq.\ (\ref{deltae2}) (circles),
and the model of  Holstein, Eq.\ (\ref{eq:theor5}) (squares).}
\end{figure}

   In Fig.~\ref{fig:invs} we show the generated distributions of events as
functions of the invariants $s_1$, $s_2$, $t_1$, and $t_2$ as obtained 
using chiral perturbation theory at ${\cal O}(p^6)$ 
[see Eq.\ (\ref{eq:theor3})] with 
the low-energy constants of Eq.\ (\ref{cestimate}) and the most prominent 
electromagnetic correction of Eq.\ (\ref{deltae2}).
   The results are based on the generation of $100\,000$ events restricted
to the kinematic region
\begin{eqnarray}
\label{kin_reg_gen}
&&2\, M_{\pi}^2<s_1<10\, M_{\pi}^2,\nonumber\\
&&4\, M_\pi^2 < s_2 < 10\, M_\pi^2,\nonumber\\
&&-0.015\,\mbox{GeV}^2 < t_1 < -0.001\,\mbox{GeV}^2,\nonumber\\
&&-0.269\, \mbox{GeV}^2<t_2<0.
\end{eqnarray}
   The regions have been chosen such as to avoid the $1/t_1$ pole
and to be far enough away from the $\rho$-meson production threshold.
   In Fig.\ \ref{fig:mv_dif} we show the differential cross section
$d\sigma{/}dt_1$ as a function of $t_1$ for three different choices
of the low-energy constants of Eq.\ (\ref{cestimate}).
   Clearly, $M_V=0.2$ GeV (triangles) would correspond to unrealistically
large higher-order terms which is supported by the drastically different
behavior for this case.
   Figure \ref{fig:mod_8_9} illustrates how the inclusion of the 
electromagnetic correction of Eq.\ (\ref{deltae2}) affects  
the differential cross section $d\sigma{/}dt_1$.
   The calculational errors in the plotted points of Figs.\  
\ref{fig:mv_dif} and \ref{fig:mod_8_9} range from 0.1 \% at 
$t_1=-0.015$ GeV$^2$ to 0.3 \% at $t_1=-0.001$ GeV$^2$.
   Finally, Fig.\ \ref{fig:mod_comp} contains a comparison of different 
(model) calculations for the  differential cross section $d\sigma{/}dt_1$.
   Here, the calculational errors in the plotted points range from
0.03 \% at $t_1=-0.015$ GeV$^2$ to 0.07 \% at $t_1=-0.001$ GeV$^2$.

\section{Conclusion} \label{section_summary}
   We have studied the reaction $\pi^- e^- \rightarrow \pi^- e^- \pi^{0}$ with
the purpose of obtaining information on the  $\gamma \pi \rightarrow \pi\pi$
anomalous amplitude ${\cal F}_{3\pi}$.
   In Table \ref{table_sigmatot} we have summarized the results of various
phenomenological models and of a full calculation at ${\cal O}(p^6)$ in chiral
perturbation theory for the total cross section using the kinematical
conditions of Ref.\ \cite{Amendolia:1985bs}.
   In particular, by integrating the ChPT results using 
Monte Carlo techniques we obtain $\sigma= 2.05 $ nb at ${\cal O}(p^6)$ and 
$\sigma= 2.17 $ nb after including the dominant electromagnetic correction.
   Both results are in good agreement with the experimental cross section of 
$\sigma= (2.11 \pm 0.47)$ nb \cite{Amendolia:1985bs}.
   On the basis of the ChPT results one would extract from the 
the experimental cross section as amplitudes ${\cal F}_{3\pi}^{(0)\rm extr}
= (9.9 \pm 1.1)$ GeV$^{-3}$ and ${\cal F}_{3\pi}^{(0)\rm extr}
= (9.6 \pm 1.1)$ GeV$^{-3}$, respectively, 
which have to be compared with the low-energy theorem 
${\cal F}_{3\pi}=e/(4\pi^2 F_\pi^3)=9.72\,\mbox{GeV}^{-3}$. 
   We emphasize the need for new data to allow comparison of
experimental and theoretical distributions and to obtain ${\cal F}_{3\pi}$ 
with smaller uncertainty.  
   In order to further support our findings and to obtain ${\cal F}_{3\pi}$ 
with a smaller uncertainty, it would be useful for future
experiments to also consider distributions such as those shown in 
Figs.\ \ref{fig:invs} - \ref{fig:mod_comp}.

\acknowledgments
   The work of T.~E.~and S.~S.~was supported by the Deutsche
Forschungsgemeinschaft (SFB 443). The work of I.~G. and M.~A.~M.~was
supported by the Israel Science Foundation. Many thanks go to R.~Tenchini for
interesting clarifications and communications regarding the work Amendolia {\em
et al.}. We also thank A.~Vainshtein for helpful comments and encouragement. 
Furthermore, S.~S.~would like to thank A.~I.~L'vov for clarifying
comments on the calculation of Appendix \ref{appendix_direct_calculation} and
J.~Gasser and J.~Gegelia for very useful comments on interpreting the ChPT
result. 

\appendix
\section{\label{appendix_scalar_products}Scalar products}
   The ten scalar products appearing in the calculation of the differential
cross section of Eq.\ (\ref{dsigma}) may be expressed in terms of the five
standard invariants of Eq.\ (\ref{five_invariants}) as \cite{Byckling}
\begin{equation}
\begin{array}{llllll}
 2\ p_{a}\cdot p_{b}&=&s-m_{a}^{2}-m_{b}^{2}, \;\;\;\;\;\;\;\;\;\;\;\;
 &2\ p_{b}\cdot p_{2}&=&s_{2}+t_{2}-t_{1}-m_{3}^{2},  \\
 2\ p_{a}\cdot p_{1}&=&m_{a}^{2}+m_{1}^{2}-t_{1}, \;\;\;\;\;\;\;\;\;\;\;\;
 &2\ p_{b}\cdot p_{3}&=&m_{b}^{2}+m_{3}^{2}-t_{2},  \\
 2\ p_{a}\cdot p_{2}&=&s_{1}+t_{1}-t_{2}-m_{1}^{2}, \;\;\;\;\;\;\;\;\;\;\;\;
 &2\ p_{1}\cdot p_{2}&=&s_{1}-m_{1}^{2}-m_{2}^{2},  \\
 2\ p_{a}\cdot p_{3}&=&s-s_{1}+t_{2}-m_{b}^{2}, \;\;\;\;\;\;\;\;\;\;\;\;
 &2\ p_{1}\cdot p_{3}&=&s-s_{1}-s_{2}+m_{2}^{2},  \\
 2\ p_{b}\cdot p_{1}&=&s-s_{2}+t_{1}-m_{a}^{2}, \;\;\;\;\;\;\;\;\;\;\;\;
 &2\ p_{2}\cdot p_{3}&=&s_{2}-m_{2}^{2}-m_{3}^{2}.
\end{array}
\label{scalar_products}
\end{equation}

\section{\label{squared_amplitude} The squared amplitude}
   In order to evaluate Eq.\ (\ref{M2overline}), one makes use of the
double epsilon expression
\begin{displaymath}
\epsilon^{\mu\nu\rho\sigma}
\epsilon^{\mu'\nu'\rho'\sigma'}
=-\mbox{det}(g^{\alpha\alpha'}),\quad
\alpha=\mu,\nu,\rho,\sigma,\quad
\alpha'=\mu',\nu',\rho',\sigma'.
\end{displaymath}
   The evaluation of the expression in brackets in Eq.\ (\ref{M2overline})
is straightforward but tedious and we only quote the final result:
\begin{eqnarray}
\label{evaluation}
\lefteqn{\left(2 p_a\cdot F\,p_1\cdot F+\frac{1}{2}q^2 F\cdot F\right)}
\nonumber\\
&=&\left(2p_a\cdot p_1+\frac{1}{2}q^2\right)
[p_b^2(p_2\cdot p_3)^2+p_2^2(p_b\cdot p_3)^2
+p_3^2(p_b\cdot p_2)^2
-p_b^2 p_2^2 p_3^2-2 p_b\cdot p_2 p_b\cdot p_3 p_2\cdot p_3]\nonumber\\
&&+2p_a\cdot p_b p_1\cdot p_b [p_2^2 p_3^2-(p_2\cdot p_3)^2]\nonumber\\
&&+2p_a\cdot p_2 p_1\cdot p_2 [p_b^2 p_3^2-(p_b\cdot p_3)^2]\nonumber\\
&&+2p_a\cdot p_3 p_1\cdot p_3 [p_b^2 p_2^2-(p_b\cdot p_2)^2]\nonumber\\
&&+2(p_a\cdot p_b p_1\cdot p_2+p_a\cdot p_2p_1\cdot p_b)
(p_b\cdot p_3 p_2\cdot p_3-p_b\cdot p_2 p_3^2)\nonumber\\
&&+2(p_a\cdot p_b p_1\cdot p_3+p_a\cdot p_3 p_1\cdot p_b)
(p_b\cdot p_2 p_2\cdot p_3-p_b\cdot p_3 p_2^2)\nonumber\\
&&+2(p_a\cdot p_2 p_1\cdot p_3+p_a\cdot p_3 p_1\cdot p_2)
(p_b\cdot p_2 p_b\cdot p_3-p_2\cdot p_3 p_b^2).
\end{eqnarray}
   We deliberately did not express the scalar products $p_b^2$,
$p_2^2$, and $p_3^2$ in terms of the masses,
because it is then straightforward to apply Eq.\ (\ref{evaluation})
for other processes involving particles with different masses, such as,
e.g., $\gamma^\ast(q)+K^-(p_b)\to \pi^0(p_2)+K^-(p_3)$.
   Note that Eq.\ (\ref{evaluation}) is manifestly symmetric under
both the exchange $p_1\leftrightarrow p_a$ and the exchange of any two
elements of $\{p_b,p_2,p_3\}$.
   Finally, using Eqs.\ (\ref{scalar_products}) the scalar products in
Eq.~(\ref{evaluation}) may be expressed in terms of the invariant variables
of Eq.~(\ref{five_invariants}).

\section{\label{appendix_direct_calculation}Direct calculation}
   Using the covariant normalization for both bosons and fermions, the
differential cross section for $e^-(p_a)+\pi^-(p_b)\to
e^-(p_1)+\pi^0(p_2)+\pi^-(p_3)$ can be written as
\begin{equation}
\label{app_dsigma} d\sigma=\frac{1}{4}\frac{1}{\sqrt{(p_a\cdot
p_b)^2-m_e^2M_\pi^2}}\frac{1}{(2\pi)^5}\, \overline{|{\cal M}|^2}\,
\delta^4(p_a+p_b-p_1-p_2-p_3)\frac{d^3 p_1}{2E_1} \frac{d^3p_2^\ast}{2E_2^\ast}
\frac{d^3 p_3^\ast}{2E_3^\ast},
\end{equation}
where we consider the final-state pions in their rest frame (denoted by $\ast$)
and the ejected electron in the laboratory frame.
   Integration with respect to $\vec{p}\,^\ast_3$ and $E_2^\ast$ yields
\begin{displaymath}
d\sigma=\frac{1}{4}\frac{1}{\sqrt{(p_a\cdot
p_b)^2-m_e^2M_\pi^2}}\frac{1}{(2\pi)^5}\, \overline{|{\cal M}|^2}\, \frac{d^3
p_1}{2E_1} \frac{|\vec{p}\,^\ast_2|d\Omega_2^\ast}{4\sqrt{s_2}}.
\end{displaymath}
   Let the pion beam define the positive $z$ axis. Using
\begin{displaymath}
\int_0^{2\pi}\frac{d^3 p_1}{E_1}=\pi dE_1 dp_{1z}
\end{displaymath}
in combination with
\begin{eqnarray*}
2|\vec{p}_b|p_{1z}&=&s_2-M_\pi^2-\frac{t_1}{2m_e^2}(m_e^2+s-M_\pi^2),\\
2 m_e E_1&=&2m_e^2-t_1,\\
m_e |\vec{p}_p|&=&\sqrt{(p_a\cdot p_b)^2-m_e^2 M_\pi^2},
\end{eqnarray*}
we obtain
\begin{displaymath}
\pi d E_1 dp_{1z}=\frac{\pi}{4}\frac{dt_1 ds_2}{\sqrt{(p_a\cdot p_b)^2-m_e^2
M_\pi^2}}
\end{displaymath}
   and can thus write
\begin{equation}
\label{app_dsigma_int}
 d\sigma=\frac{1}{4}\frac{1}{\sqrt{(p_a\cdot
p_b)^2-m_e^2M_\pi^2}}\frac{1}{(2\pi)^5}\, \overline{|{\cal
M}|^2}\, \frac{\pi dt_1 ds_2 }{4\sqrt{(p_a\cdot
p_b)^2-m_e^2M_\pi^2}}
\frac{|\vec{p}\,^\ast_2|d\Omega_2^\ast}{4\sqrt{s_2}}.
\end{equation}
   To obtain the kinematic boundaries of $s_2$ and $t_1$ one can either solve
for $t_1$ in terms of $s_2$, or vice versa \cite{Byckling},
\begin{eqnarray}
t_1^\pm&=&m_a^2+m_1^2-\frac{1}{2s}[(s+m_a^2-m_b^2)(s-s_2+m_1^2)\mp
\lambda^{1/2}(s,m_a^2,m_b^2)\lambda^{1/2}(s,s_2,m_1^2)],\nonumber\\
s_2^\pm&=&s+m_1^2-\frac{1}{2m_a^2}[(s+m_a^2-m_b^2)(m_a^2+m_1^2-t_1) \mp
\lambda^{1/2}(s,m_a^2,m_b^2)\lambda^{1/2}(t_1,m_a^2,m_1^2)],
\nonumber\\ \label{kinbound}
\end{eqnarray}
where $\lambda(x,y,z)$ is defined in Eq.\ (\ref{lambda}).

   In order to test our numerical integration programs, we
evaluated Eq.\ (\ref{app_dsigma_int}) under the following simplifying 
assumptions:
We neglected terms containing the square of the electron mass and we assumed
${\cal F}_{3\pi}$ to be constant [see Eq.\ (\ref{eq:int5})].
   Under these assumptions we obtain for the angular integral
   \begin{equation}
   \label{angular_integral}
   \int d\Omega_2^\ast(4p_a\cdot F p_1\cdot F+t_1 F\cdot
   F)=-\frac{\pi}{6}t_1(s_2-4M_\pi^2)[t_1^2+2 t_1(s-s_2-M_\pi^2)
   +(s-M_\pi^2)^2+(s-s_2)^2].
\end{equation}
   Inserting Eq.\ (\ref{angular_integral}) into Eq.\ (\ref{app_dsigma_int})
and using Eq.\ (\ref{M2overline}) results in
\begin{equation}
\label{app_dsigmadt1ds2}
 \frac{d\sigma}{dt_1 ds_2}=
-\frac{1}{6144\pi^3}\frac{1}{(s-M_\pi^2)^2}|e{\cal F}_{3\pi}|^2 \frac{(s_2-4
M_\pi^2)^{3/2}}{\sqrt{s_2}t_1}[t_1^2+2t_1(s-s_2-M_\pi^2)
+(s-M_\pi^2)^2+(s-s_2)^2]
\end{equation}
for $m_e^2\to 0$ and ${\cal F}_{3\pi}=\mbox{const}$.
   We first integrate Eq.\ (\ref{app_dsigmadt1ds2}) with respect to $t_1$ from
$t_1^-=-(s-M_\pi^2)(s-s_2)/s$ to $t_1^+=t_1^c$, where $t_1^c$ is an 
experimental cut on the maximal value of $t_1$:
\begin{eqnarray}
\label{app_dsigmads2}
\int_{t_1^-}^{t_1^c}dt_1\frac{d\sigma}{dt_1ds_2}&=&\frac{1}{6144
\pi^3}\frac{1}{(s-M_\pi^2)^2}|e{\cal
F}_{3\pi}|^2\frac{(s_2-4M_\pi^2)^{3/2}}{\sqrt{s_2}}\nonumber\\
&&\times [\frac{1}{2}
({t_1^-}^2-{t_1^c}^2)+2(s-s_2-M_\pi^2)(t_1^--t_1^c)+[(s-M_\pi^2)^2+(s-s_2)^2]
\ln\left(\frac{t_1^-}{t_1^c}\right)],\nonumber\\
\end{eqnarray}
where $4M_\pi^2\leq s_2\leq s_2^c=s[1+t_1^c /(s-M_\pi^2)]$ as implied by the
physical boundary conditions.
   A numerical integration of
\begin{equation}
\sigma(t_1^c)=\int_{4 M_\pi^2}^{s_2^c} ds_2 \int_{t_1^-}^{t_1^c} dt_1
\frac{d\sigma}{dt_1 ds_2}
 \label{app_tot_sigma}
\end{equation}
yields $\sigma(t_1^c)=1.864$ nb for $t_1^c=-0.001$ GeV$^2$ which agrees within
less than 0.5 \% with the Monte Carlo result 1.855 nb (we used $M_\pi=139.57$
MeV).

\begin{table}
\caption{\label{table_sigmatot}Total cross section for
$\pi^- e^- \rightarrow \pi^- e^- \pi^{0}$ as obtained in different models and
chiral perturbation theory (see text).
   The third column contains the physical threshold amplitudes for $q^2=0$.
   In the last column we made use of the experimental result 
$(2.11 \pm 0.47)$~nb of Ref.\ \cite{Amendolia:1985bs} to convert this into 
the value for ${\cal F}_{3\pi}^{(0)\rm extr}$ which would be extracted based 
on the given model: 
${\cal F}_{3\pi}^{(0)\rm extr}=9.72\,
\sqrt{\sigma_{\rm exp}/\sigma_{\rm model}}$ GeV$^{-3}$.
   Here we follow the common practice to quote the extracted values.
   The error only reflects the experimental error and does not include
any error estimate implied by the models.  
   We stress that in the framework of ChPT the overall factor of 
Eq.\ (\ref{eq:theor3}) is not a free parameter. 
   Therefore, the ``extracted'' values for the ChPT calculation 
have to be taken with a grain of salt (see the discussion in the text).} 
\vspace{1em}
\begin{tabular}{lccc}
\hline \hline
&&\\
Model/theory & cross section & ${\cal F}_{3 \pi}^{\rm thr}$ & 
${\cal F}_{3\pi}^{(0)\rm extr}$\\
& [nb] & [GeV$^{-3}$] & [GeV$^{-3}$]\\
&&&\\
1) 
${\cal F}_{3\pi}=\frac{e}{4\pi^2 F_\pi^3}=9.72\,\mbox{GeV}^{-3}$ 
& $1.92$ 
& $9.7$ 
& $10.2 \pm 1.1$\\
2) Terent'ev, Eq.\ (\ref{eq:theor1}) with $\Delta_\rho=0.5$ 
and $\Delta_\omega=0$
& $2.80$ 
& $10.3$ 
& $8.4 \pm 0.9$\\
3) Terent'ev, Eq.\ (\ref{eq:theor1}) with $\Delta_\rho=0.5$ and
$\Delta_\omega=1.5$
& $2.62$ 
& $10.3$
&$8.7 \pm 1.0$\\
4) Terent'ev, Eq.\ (\ref{eq:theor1}) with $\Delta_\rho=0.35$ and
$\Delta_\omega=0$ 
&$2.51$ 
&$10.1$ 
&$8.9 \pm 1.0$ \\
5) Terent'ev, Eq.\ (\ref{eq:theor1}) with $\Delta_\rho=0.35$ and
$\Delta_\omega=3.2$
& $2.18$ 
& $10.1$
&$9.6 \pm 1.1$\\
6) Rudaz, Eq.\ (\ref{eq:theorudaz})
& $2.36$ 
& $10.0$
&$9.2\pm 1.0$ \\
7) ChPT at ${\cal O}(p^6)$ [Eq.\ (\ref{eq:theor3})] without $q^2$ dependence  
& $2.33$ 
& $10.4$
&$9.2\pm 1.0$\\
8) ChPT at ${\cal O}(p^6)$ [Eq.\ (\ref{eq:theor3})] with $q^2$ dependence 
& $2.05$
& $10.4$
& $9.9\pm 1.1$\\
9) ChPT at ${\cal O}(p^6)$ [Eq.\ (\ref{eq:theor3})] with $q^2$ dependence 
&&&\\
plus electromagnetic correction of Eq.\ (\ref{deltae2})
&$2.17$ 
&$12.1$
&$9.6\pm 1.1$\\
10) ChPT at ${\cal O}(p^6)$ with modified 
dependence of Eq.\ (\ref{replacement})
& $2.83$ 
& $10.5$
&$8.4\pm 0.9$\\
11) Holstein, Eq.\ (\ref{eq:theor5}) 
& $3.05$ 
& $10.4$
& $8.1\pm 0.9$\\
\hline
\end{tabular}
\end{table}

\end{document}